\documentclass[aps,prl,twocolumn,amsmath,amssymb,floatfix,longbibliography]{revtex4-2}

\usepackage{color}
\usepackage{mathrsfs}
\usepackage{graphicx}
\usepackage{dcolumn}
\usepackage{bm}
\usepackage{hyperref}
\usepackage{amsmath,amsfonts,amssymb,ulem}
\usepackage{epstopdf}
\usepackage{xcolor}

\newcommand{\re}[1]{{\color{black}#1}}

\usepackage{graphicx}

\usepackage{upgreek}
\newcommand{\micron}{{\upmu\mathrm{m}}}
\begin{document}

\title{Energy-chirp compensation of laser-driven ion beams enabled by structured targets}
\author{Zheng Gong$^{1,2}$}\email[]{Current address: Max-Planck-Institut f\"{u}r Kernphysik, Saupfercheckweg 1, 69117 Heidelberg, Germany}
\author{Stepan S. Bulanov$^3$}
\author{Toma Toncian$^4$}
\author{Alexey Arefiev$^5$} 

\affiliation{$^1$Center for High Energy Density Science, The University of Texas at Austin, Austin, TX 78712, USA}
\affiliation{$^2$SKLNPT, KLHEDP, CAPT and School of Physics, Peking University, Beijing 100871, China}
\affiliation{$^3$Lawrence Berkeley National Laboratory, Berkeley, California 94720, USA}
\affiliation{$^4$Institute for Radiation Physics, Helmholtz-Zentrum Dresden-Rossendorf e.V., 01328 Dresden, Germany}
\affiliation{$^5$University of California at San Diego, La Jolla, CA 92093, USA}

\begin{abstract} 
We show using 3D simulations that the challenge of generating dense mono-energetic laser-driven ion beams with low angular divergence can be overcome by utilizing structured targets with a relativistically transparent channel and an overdense wall. In contrast to a uniform target 
that produces a chirped ion beam, the target structure facilitates formation of a dense electron bunch whose longitudinal electric field reverses the energy chirp. This approach works in conjunction with existing acceleration mechanisms, augmenting the ion spectra. For example, our 3D simulations predict a significant improvement  for a 2 PW laser pulse with a peak intensity of $5 \times 10^{22}$ W/cm$^2$. The simulations show a mono-energetic proton peak in a highly desirable energy range of 200~MeV with an unprecedented charge of several nC and relatively low divergence that is below 10$^{\circ}$.
\end{abstract}

\maketitle

Development of laser-driven ion accelerators attracts considerable attention since the generated ion beams are essential to various cross-disciplinary applications~\cite{Mourou_etal_2006, Daido_review_ion, macchi_2013_RMP, bulanov2014laser}. The state-of-the-art high-power laser systems~\cite{danson2019petawatt_review,ELI_NP,yoon2021realization} have the potential to enable compact and efficient high repetition rate ion accelerators. Depending on the interaction parameters, the ion acceleration can follow one of multiple scenarios~\cite{bulanov.pop.2016}. 
The most studied and used scenario is the Target Normal Sheath Acceleration (TNSA)~\cite{wilks.pop.2001,mora2003plasma_TNSA,roth.lpia.2013,wagner2016maximum_85MeV}, whereas the most efficient one in terms of energy transfer from the laser to the ions is the Radiation Pressure Acceleration (RPA)~\cite{esirkepov2004highly, yan2008generating, macchi2009light, qiao2009stable,kim2016radiation_93MeV}. Successful realization of RPA requires ultra-thin targets that stay intact during the interaction, but this can be hard to achieve due to laser pre-pulse and plasma instabilities~\cite{kaluza2004influence, mckenna2008effects, steinke2009efficient, pegoraro2007photon, dollar.prl.2012, sgattoni2015laser_RTI, wan2016physical_RTI, wan2020physical_I}. These difficulties can be circumvented by employing acceleration mechanisms such as shockwave acceleration \cite{silva2004proton_shock, haberberger2012collisionless, zhang2017collisionless}, hole-boring radiation pressure acceleration \cite{sentoku2003high, robinson2009relativistically, naumova2009hole}, and magnetic vortex acceleration (MVA) \cite{kuznetsov.ppr.2001, fukuda2009energy, bulanov.pop.2010, bulanov.prstab.2015, nakamura2010high_MVA, helle2016laser, sharma.scirep.2018,park.pop.2019} that involve a thick plasma with a near-critical electron density, i.e. the density close to  \re{$n_{cr} \equiv m_e \omega^2 / 4 \pi e^2\approx 1.8\times 10^{21}$ cm$^{-3}$} for a $800$ nm wavelength electromagnetic wave \cite{Mourou_etal_2006,esarey.rmp.2009}. Here $\omega$ is the carrier frequency of the laser pulse and $m_e$ ($e$) is the electron mass (charge).

\begin{figure}[tb]
\includegraphics[width=0.95\columnwidth]{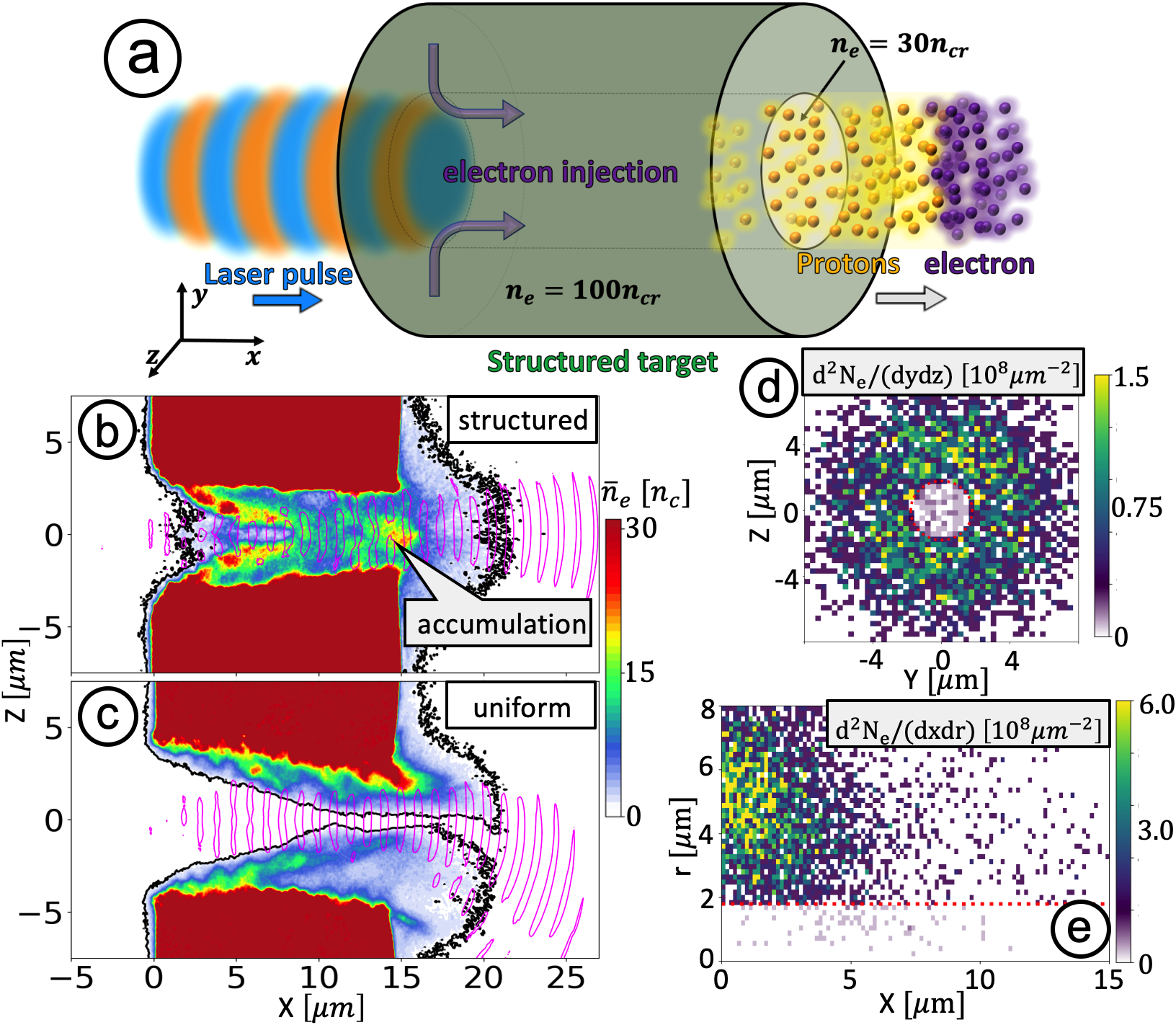}
\caption{(a) Schematic setup. (b) and (c) Time-averaged electron densities $\overline{n}_e$ in the $(x,z)$-plane at $y=0$ for structured and uniform targets. Black and magenta contours correspond to $\overline{n}_e = n_{cr}$ and $a_0=60$. (d) and (e) Initial location of the electrons from the bunch located in (b) at $12~\micron < x < 16~\micron$, $r < 2~\micron$ and marked as ``accumulation''. The dotted red line is the initial location of channel wall.} \label{fig_scheme}
\end{figure} 


Most applications have requirements for individual particle energies and their collective charge~\cite{Mourou_etal_2006, Daido_review_ion, macchi_2013_RMP, bulanov2014laser}. \re{Meeting both requirements has proven to be extremely challenging for energies in excess of 100~MeV per nucleon. Achieving a required number of particles in a specific energy range is considered a grand challenge and a prerequisite for wide utilization of high-power laser systems for a number of applications.} For example, TNSA and MVA generate broad ion energy spectra that can contain very energetic ions~\cite{wagner2016maximum_85MeV}. 
However, the ion numbers decrease with energy, so there are very few of these ions, meaning that merely reaching high energies is insufficient for these mechanisms. Recent progress in target fabrication has opened up to experimental research 
\re{novel target configurations~\cite{snyder2019relativistic, channel_ion_acc_PRE2020,rinderknecht2021relativistically}} and reinvigorated the study of laser-driven ion acceleration. These configurations have the potential to relax what was previously perceived as fundamental constraints, e.g. the inability to produce mono-energetic peaks for certain otherwise promising acceleration mechanisms such as MVA.

In this Letter, we show using 3D kinetic particle-in-cell (PIC) simulations how a proton acceleration mechanism that generates a broad spectrum with a monotonically decreasing number of energetic particles can be augmented to achieve a mono-energetic peak. A new physics phenomenon is the emergence of a dense forward-moving electron bunch enabled by a structured target with a pre-filled channel shown in Fig.~\ref{fig_scheme}a and a strong plasma magnetic field. 
Prior to the arrival of the bunch, the ions have a positive chirp with the ion energy increasing along the axis. \re{The bunch generates a strong positive longitudinal electric field, so, as it catches up with the energetic ions, this field compensates the chirp by accelerating the lower energy ions and produces a mono-energetic peak in the ion spectrum.} The electron bunch has also a focusing transverse electric field that reduces the divergence of the accelerated protons. 
\re{We refer to the described mechanism as the laser ion-shotgun acceleration (LISA), because the channel is akin to a barrel of a shotgun that builds up an expulsion force and propels a bullet - the collimated proton beam in our case.} \re{We note that the structured  targets, which enable LISA, have evolved from being a concept~\cite{stark.prl.2016} to a commercial product used in experiments~\cite{rinderknecht2021relativistically} in the period of just six years.}


\begin{figure*}[tbp] 
\includegraphics[width=1.98\columnwidth]{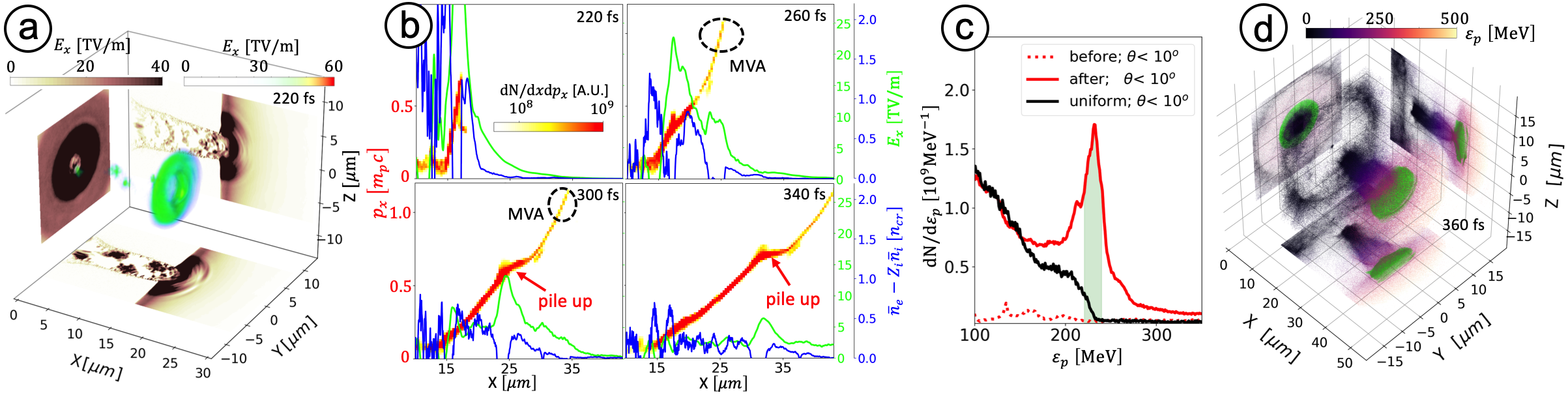}
\caption{(a) Longitudinal electric field $E_x$ where the projections exhibit the distribution at the slice of $x=16 \mu m$, $y=0$ and $z=0$. (b) Proton distribution in $x-p_x$ space, where the green line profiles the field $E_x$ and the blue line denotes the net charge density $\overline{n}_e-Z_i\overline{n}_i$. (c) Proton energy spectra. (d) The distribution of the generated proton beams, where the lime dots represent the protons within energy range of $220<\varepsilon_p[\mathrm{MeV}]<240$ [shadow green area in (d)].} \label{fig_accelerating}
\end{figure*}

\begin{figure}[tb]
\includegraphics[width=0.99\columnwidth]{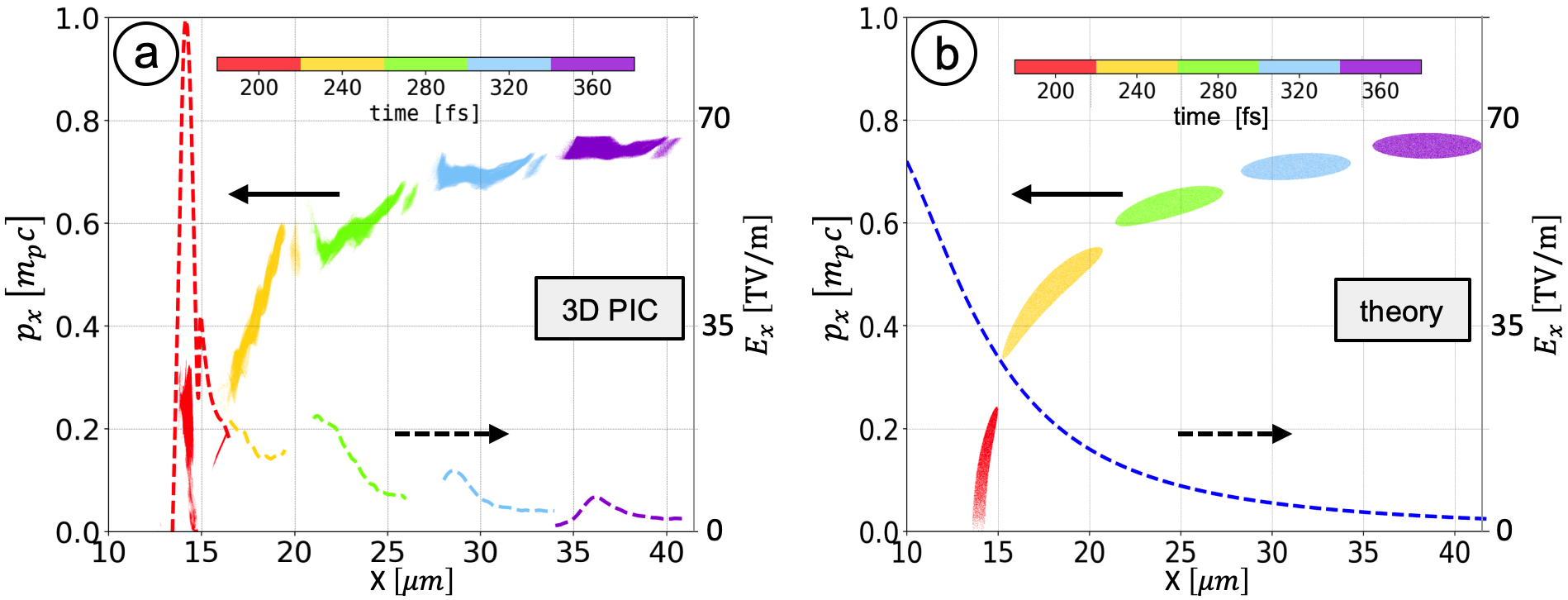}
\caption{(a) Proton distribution in $(x,p_x)$ space, where the colors correspond to different time and dashed lines denote the field $E_x$ imposed on the proton bunch. (b) is same as (a) but for the theoretically predicted proton evolution, where the blue dashed line refers to the estimated electric field.} \label{fig_chirp_comp}
\end{figure} 

\begin{figure*}[tb]
\includegraphics[width=1.95\columnwidth]{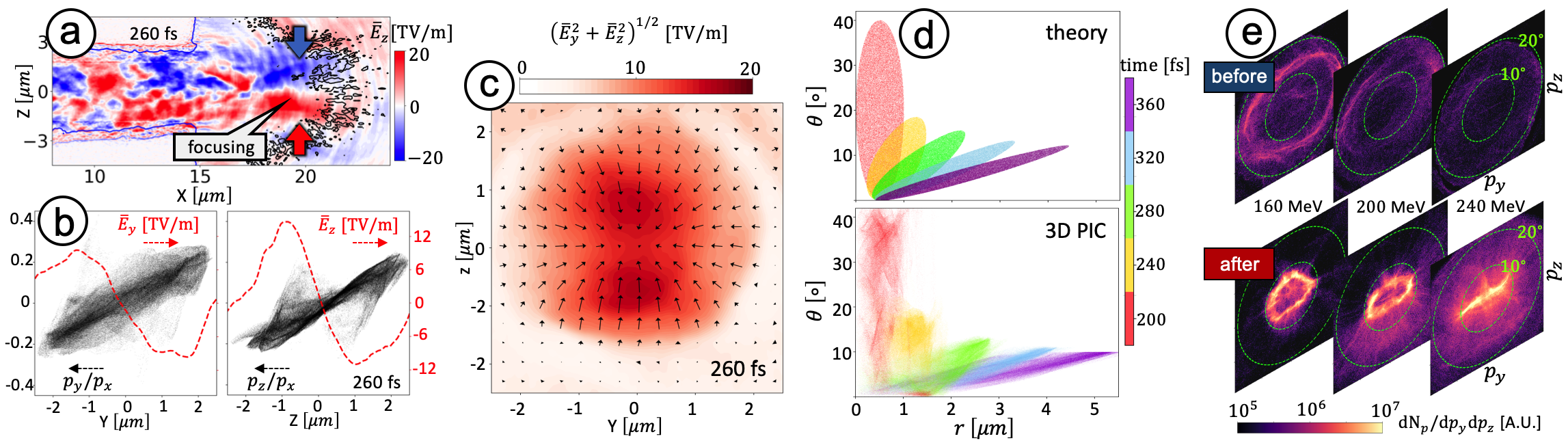}
\caption{(a) Transverse electric field $\overline{E}_z$ at the slice of $y=0$. (b) Distribution of protons in transverse $p_y-y$ ($p_z-z$) space. (c) Radial focusing electric field $\overline{E}_r=(\overline{E}_y^2+\overline{E}_z^2)^{1/2}$ averaged over $18<x[\mu m]<21$ where black arrows denote the field direction. (d) Time evolution of the protons within the monoenergetic part $220<\varepsilon_p[\mathrm{MeV}]<240$ in $r-\theta$ space. (e) Proton angular distribution \textit{before} and \textit{after} the transversely focusing process.} \label{fig_collimation}
\end{figure*} 

\textit{Structured vs uniform target} --- \re{We have performed two 3D PIC simulations using the fully-relativistic code EPOCH~\cite{arber2015contemporary}}: a simulation with the structured target from Fig.~\ref{fig_scheme}a and a simulation with a uniform target with the density equal to the density of the channel. Both simulations use the same linearly polarized 2~PW laser (\re{$\lambda_0$=$1\micron$} wavelength) focused at the target surface ($x = 0$). In the absence of the target, transverse laser electric field in the focal plane is \re{$E_y = E_0\exp{(-r^2/w_0^2)}\exp{[-4(c t - x)^2/(c\tau)^2]}\cos(\omega t)$}, where $w_0 = 2.2~\micron$ and $\tau=120$~fs. The normalized laser amplitude $a_0 \equiv |e| E_0 / m_e c \omega \approx 190$ corresponds to a peak laser intensity of $5 \times 10^{22}$~W/cm$^2$, where $c$ is the speed of light. The uniform target is a hydrogenic plasma with electron density $n_e = 30 n_{cr}$. The structured target has a cylindrical channel of radius \re{$R_{ch}=1.8~\micron$} that is filled with the same plasma. The bulk is a carbon plasma with $n_e = 100 n_{cr}$. Both targets are $L_T=15~\micron$ thick. We use 50 (10) macro-particles per cell to represent ions (electrons). The simulation domain is \re{$60~\micron \times 30~\micron \times 30~\micron$ with $1200 \times 360 \times 360$ cells}.

As the high intensity laser starts to interact with these targets, it quickly ionizes them. Then the laser is guided through these targets, where either a self-generated channel, Fig.~\ref{fig_scheme}c, or the pre-manufactured channel, Fig.~\ref{fig_scheme}b, facilitate this. Since the thickness of these targets is rather small, they are transparent to the relativistically intense laser pulses, i.e., the thickness is smaller than the depletion length of the laser pulse in the plasma of such density \cite{bulanov.pop.2010,bulanov.prstab.2015,park.pop.2019}, $L_T<L_D=(c\tau K)(a_{ch}n_{cr})/n_e$, which is around 40 $\mu$m for given laser and plasma parameters. Here $K=0.074$ is geometrical factor from depletion length calculation \cite{bulanov.pop.2010,bulanov.prstab.2015,park.pop.2019} and $a_{ch}$ is the field amplitude inside the plasma channel. As the laser pulse exits the target, a proton acceleration regime similar to the MVA  takes place in both cases, with a cutoff energy of about 450 MeV 

However, there are important distinctions between these two targets. First, a structured target provides a more stable propagation of an intense laser pulse due to high density walls (compare Fig.~\ref{fig_scheme}b and Fig.~\ref{fig_scheme}c). Second, during the interaction a forward moving electron density maximum is formed near the exit of the channel (see \ref{fig_scheme}b), which is absent in the other case. Such electron density evolution leads to the formation of strong forward moving focusing and accelerating for protons electric fields. This results in the sharp peak in the proton spectrum around 220 MeV, see Fig.~\ref{fig_accelerating}c. This is exactly the MVA spectrum augmentation, which was mentioned above.

In what follows we describe in detail the laser pulse interaction with a structured target and discuss the underlying physical mechanisms, which generate high charge well collimated proton bunch. 

\textit{Electron extraction} --- When the laser transverse electric field, $E_y$, which is imposed on the front surface of the target bulk, the electrons originating from the bulk inner edge are prone to be extracted by the field $E_y$. Then, these injected electrons with an imprinted transverse momentum $p_y\sim |e|E_y/(m_ec\omega)$ are deflected by the magnetic force $(\boldsymbol{p}_y/\gamma)\times \boldsymbol{B}$ to move forward. The density of these electrons can be estimated from requiring the balance between the energy gain in the ponderomotive potential and in the field of the ion core in the channel \cite{bulanov.pop.2010, bulanov.prstab.2015, park.pop.2019}. Then, we obtain:
\begin{equation} \label{eq:density}
    n_e/n_{cr}=(2/K)^{1/2}(\lambda/\pi R_{ch})^3 (P/P_c)^{1/2},
\end{equation}
where $P_c=2 m_e^2/e^2=17$ GW. Then, the magnetic field, generated by these electrons is of the order of $eB/m\omega c=r n_e/\lambda n_{cr}$ for $r<R_{ch}$, or from Eq. (\ref{eq:density}):
\begin{equation}
    eB/m\omega c=(2/K)^{1/2}(r\lambda^2/\pi^3 R_{ch}^3) (P/P_c)^{1/2}.
\end{equation}
The maximum value of this field at $r=R_{ch}$ is 
\begin{equation}
    B_{max}=0.12 E_0\lambda/R_{ch},
\end{equation}
where $E_0$ is the laser peak field. In Fig.~\ref{fig_scheme}d-e, the distribution of the initial position of electrons (marked by "accumulation" in Fig.~\ref{fig_scheme}b) demonstrates that electrons, which maintain the strong electric field, come from the front side of target bulk region. Therefore, the merits of channel structure is not only guiding the light propagation but also providing sufficient electron population to generate the strong longitudinal electric field for proton acceleration (see Fig.~\ref{fig_accelerating}a).  

\textit{Energy chirp compensation} --- The distribution of protons within the divergence angle $\theta=\arctan(p_r/p_x)<10^\circ$ in $(x,p_x)$ space is shown in Fig.~\ref{fig_accelerating}b, where $p_r=(p_y^2+p_z^2)^{1/2}$. The field $E_x$ actively piles up a dense bunch proton with the momentum $p_x\sim 0.7m_pc$ and kinetic energy $\varepsilon_p\sim 200$ MeV at $t=340$ fs. After interaction with the moving $E_x$, this piled-up proton bunch has a monoenergetic feature in energy spectrum, as red solid line in Fig.~\ref{fig_accelerating}c, where the spectrum (red dotted line) before the interaction at $t=240$ fs is also plotted for comparison. Here, the proton energy spectrum of the case with the same parameters but a uniform target is profiled as well to certify that the monoenergetic proton bunch is a unique feature for acceleration mechanism involved with the structured channel target. The spatial distribution of the protons in Fig.~\ref{fig_accelerating}d indicates that an energetic dense proton beam is produced, where the lime dots represent the protons within the energy range of $220-240$ MeV (shadow green area in Fig.~\ref{fig_accelerating}c).

For figuring out how the proton bunch is spontaneously piled up by the moving field $E_x$, we perform a detailed particle tracking of the protons nearby the peak energy of spectrum (shadow green area in Fig.~\ref{fig_accelerating}c). The proton distribution in $x-p_x$ space for time from $t=200$ to $360$ fs (see Fig.~\ref{fig_chirp_comp}a) indicates that the proton bunch with a negative energy chirp [$-\partial \mathcal{E}_p(x)/\partial x<0$] experiences a longitudinally descending accelerating field [$\partial E_x(x)/\partial x<0$]. The stronger $E_x$ is imposed on the protons with a lower energy, which results in the elimination of the proton energy chirp. Specifically, the relative energy spread is reduced from almost $100\%$ to $8.7\%$ at the end of the chirp compensation process. In order to understand the dynamics of protons, we approximate the amplitude of the moving field as $E_x(x)=(40\tau/T_0)(m_ec\omega/|e|)/\{(x/\lambda_0-8)^2+\tau/T_0\}$, by fitting the evolution of the field that protons experience from PIC simulation results. Here $T_0$ is the period of the EM field of the laser. This approximation proves to be rather accurate as it can be seen from Fig.~\ref{fig_chirp_comp}b, where the evolution of the proton bunch phase space obtained from PIC simulation and from the approximation are shown. 

\textit{Transverse focusing}---After the laser reaches the rear side and \re{exits} the target, the quasi static transverse electric field (Fig.~\ref{fig_collimation}a) induced by the forward electron flow acts as a focusing lens collimating the accelerated protons. The outward proton momentum ($p_y$ and $p_z$) is gradually counteracted by the focusing electric fields $\overline{E}_y$ and $\overline{E}_z$ (Fig.~\ref{fig_collimation}b), \re{where the bar denotes time-averaging over two laser periods.} The inward radial electric field $\overline{E}_r$ (Fig.~\ref{fig_collimation}c) is $\sim 10$ TV/m, so it reduces the proton divergence angle by $15^\circ$ within 100 fs. We approximate the radial electric field acting on the protons by $\boldsymbol{\overline{E}}_r = -\kappa_1 r\hat{e}_r$, \re{where $\kappa_1 > 0$ is a constant.} 
The transverse proton motion in this field evolves according to equation
\begin{eqnarray}
\frac{d^2r}{dt^2}+\frac{|e|\kappa_1}{m_p\gamma}r+\frac{p_r}{m_p\gamma^2}\frac{d\gamma}{dt}=0 \label{eq:basic_transverse}
\end{eqnarray}
obtained by combining $dr/dt=p_r/(\gamma m_p)$ and $dp_r/dt=-|e|\kappa_1r$. On account of $\gamma\sim 1$ and the third term on the left-hand side of Eq.~\eqref{eq:basic_transverse} being negligible compared to the second term, the solution of Eq.~\eqref{eq:basic_transverse} reads $[r(t),p_r(t)]^\mathsf{T}=\mathcal{M}_f[r(t-t_0),p_r(t-t_0)]^\mathsf{T}$. The focusing effect of the field $E_r$ on protons is incorporated into the transmission matrix
\begin{eqnarray}\label{eq:matrix}
\mathcal{M}_f={\left\{ \begin{array}{cc}
\cos[\Omega_k(t-t_0)] & \frac{1}{\Omega_k}\sin[\Omega_k(t-t_0)] \\
-\Omega_k\sin[\Omega_k(t-t_0)] & \cos[\Omega_k(t-t_0)]
\end{array}
\right \}},
\end{eqnarray}
where $\Omega_k=\sqrt{|e|\kappa_1/m_p\gamma}$ and $t_0$ is the initial time.

In the focusing field, the protons with large divergence $\theta\sim 30^\circ$ are gradually collimated to a cone of $\theta\sim 10^\circ$ (lower panel of Fig.~\ref{fig_collimation}d). If a bunch of protons is initialized within an ellipse in $(r,\theta)$ space, the theoretically predicted proton evolution (upper panel of Fig.~\ref{fig_collimation}d) based on transmission matrix $M_f$ with $\kappa_1=3.6\pi m_ec^2/(|e|\lambda^2)\approx 5.8\times10^{18}~\mathrm{V/m^2}$ in Eq.~\eqref{eq:matrix} is in reasonable agreement with the PIC simulation results. The collimation effect can also be seen by comparing the proton distribution in the transverse momentum plane $(p_y,p_z)$ prior to (or after) the arrival of the forward moving electron population (Fig.~\ref{fig_collimation}e). After the transverse focusing, a strong pinching effect leads to the proton concentrating in a cone of $\theta<10^\circ$, whereas the distribution has a ring-type structure that is rather broad, with a characteristic opening angle of $\theta\approx 20^\circ$ before the focusing.

\begin{figure}[tb]
\includegraphics[width=0.98\columnwidth]{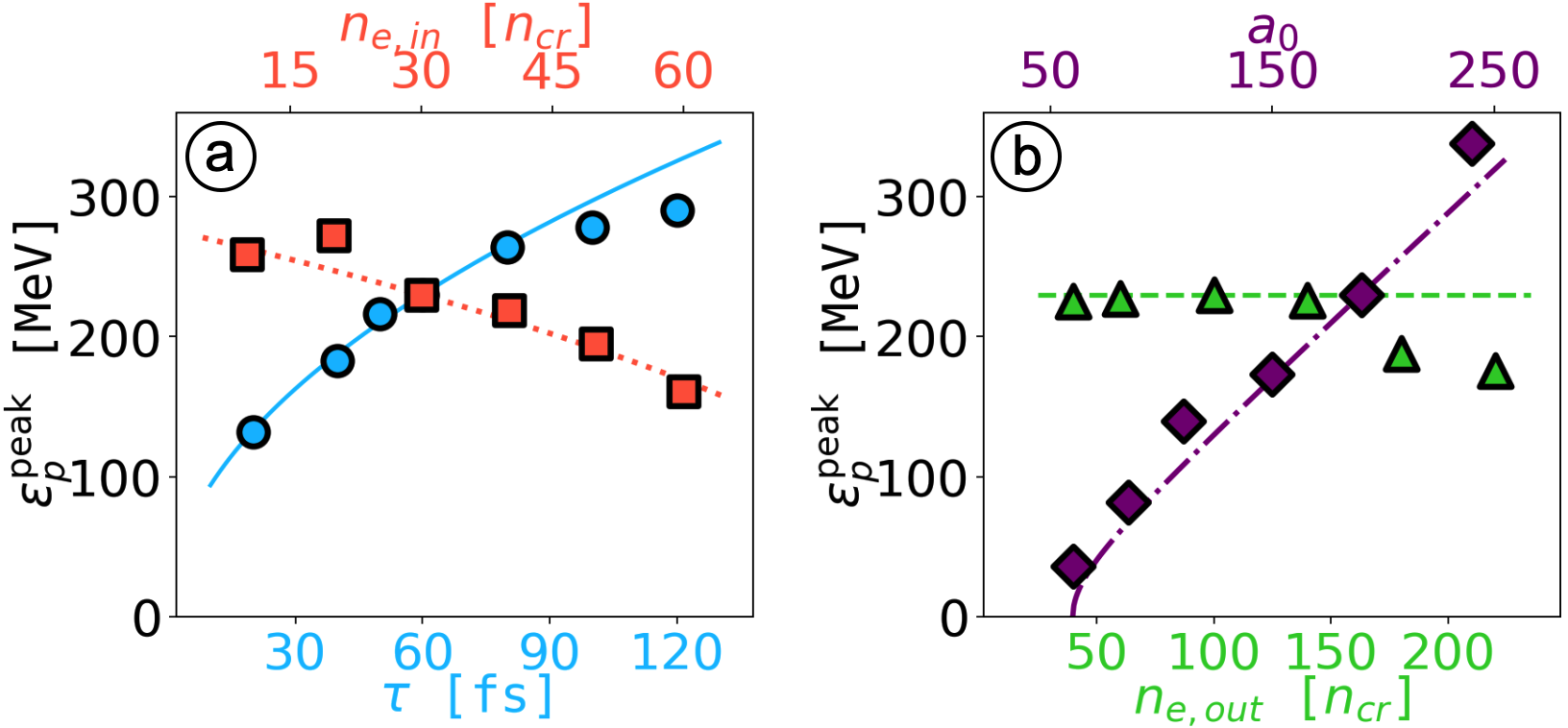}
\caption{\re{Estimates (curves: blue solid, red dotted, green dashed, purple dot-dashed) for peak energies $\varepsilon_p^\mathrm{peak}$ together with 3D PIC results (circles, squares, triangles, and diamonds) are shown for different values of (a) laser pulse duration $\tau$ and channel plasma density $n_{e,in}$, and (b) wall density $n_{e,out}$ and normalized laser field amplitude $a_0$. The range of $a_0$ corresponds to a range of laser power from 0.15 to 3.5 PW.}} \label{fig_en_scaling}
\end{figure} 

\textit{Discussion} --- \re{The presented mechanism has two distinct stages. The proton energy gain primarily happens during the initial stage that is present for both uniform and structured targets and is characterized by a broad proton spectrum with a sharp cut-off. The energy chirp-compensation and focusing occur during the second stage that is only present when using structured targets.}


\re{We have performed four scans, shown in Fig.~\ref{fig_en_scaling}, to establish the dependence of the mono-energetic peak's energy $\varepsilon_p^\mathrm{peak}$ on target and laser parameters. Each marker represents a separate 3D PIC simulation. In each scan, only a single parameter is varied while the other parameters remain the same as in the original simulation presented earlier. We find that $\varepsilon_p^\mathrm{peak}$ increases with both the normalized laser amplitude $a_0$ and with the laser pulse duration $\tau$. The increase with $\tau$ is particularly noteworthy, because it indicates there is an advantage of using such longer laser pulses as the 150~fs pulse of the L4 laser at ELI Beamlines~\cite{danson2019petawatt_review}. The scans over the channel, $n_{e,in}$, and bulk, $n_{e,out}$, densities shows that $\varepsilon_p^\mathrm{peak}$ is not particularly sensitive to these parameters over the explored range. In this study, we have deliberately explored the channel densities feasible for commercially available targets~\cite{rinderknecht2021relativistically}. Even though the channel densities are well-controlled, the observed lack of sensitivity is an important safeguard against possible fluctuations that can occur during target fabrication. Finally, it is important to mention that the proton charge associated with the mono-energetic peak (defined by the condition $|\varepsilon_p-\varepsilon_p^\mathrm{peak}| <20~\mathrm{MeV}$) remains relatively high in all four scans, varying between 1~nC and 12~nC.}


\textit{Summary and conclusions} --- \re{We have presented a novel mechanism of laser-driven ion acceleration that can generate a mono-energetic proton peak in a highly desirable energy range of 200~MeV with an unprecedented charge of several nC and relatively low divergence that is below 10$^{\circ}$. The mechanism is enabled by a novel target design that has been recently employed in experiments with high-power high-intensity lasers~\cite{rinderknecht2021relativistically}. The laser power and intensity used in our study are either within reach or are already available at multi-PW laser facilities such as ELI-NP, CoReLS, and ZEUS~\cite{danson2019petawatt_review,nees.cleo.2020}. Due to the combination of these factors, this work is an essential step towards solving the critical problem of generating high-charge proton beams with energies in excess of 100~MeV. 

Our approach also opens up a new avenue, associated with structured targets, within an established area of laser-driven ion acceleration.} \re{The use of these targets can potentially relax the requirements on the laser contrast and allow high repetition rate operation. The latter can be enabled by the use of a tape drive similar to the one discussed in \cite{steinke.prab.2020}.} 


\re{The pre-filled channel is essential for stabilization of laser propagation and for generation of strong transverse and longitudinal drift electric fields by pinched laser-accelerated electrons. As the field moves with the pinched electron current through an expanding energetic proton cloud at the back of the target, it focuses protons into a well-defined collimated quasi-monoenergetic beam. We present an example, where a 2~PW 120~fs laser can produce a quasi-monoenergetic peak centered at 230 MeV via our mechanism. The number of protons in a 40~MeV energy interval is about $5\times 10^{10}$ particles, which is equivalent to a charge of 7.8 nC. These parameters are relevant to various applications that require high energy, high repetition rate, high charge, quasi-monoenergetic proton sources. }

This research was supported under the National Science Foundation  – Czech Science Foundation partnership by NSF award PHY-2206777 and by AFOSR (Cont. No. FA9550-17-1-0382). SSB acknowledges support from the U.S. Department of Energy Office of Science Offices of High Energy Physics and Fusion Energy Sciences (through LaserNetUS), under Contract No. DE-AC02-05CH11231. Simulations were performed using EPOCH~\cite{arber2015contemporary}, developed under UK EPSRC grants EP/G054940, EP/G055165 and EP/G056803. HPC resources were provided by TACC. 

%

\end{document}